\documentclass[12pt]{article}
\usepackage[svgnames]{xcolor}

\newcommand{\beq}{\begin{equation}}
\newcommand{\eeq}{\end{equation}}
\newcommand{\beqa}{\begin{eqnarray}}
\newcommand{\eeqa}{\end{eqnarray}}
\newcommand{\beqar}{\begin{eqnarray*}}
\newcommand{\eeqar}{\end{eqnarray*}}

\newcommand{\eps}{\epsilon}

\newcommand{\inn}{\!\cdot\!}

\newcommand{\labell}[1]{\label{#1}} %{\label{#1}} %
\newcommand{\reef}[1]{(\ref{#1})}
\newcommand\prt{\partial}

\newcommand\cF{{\cal F}}

\newcommand\cM{{\cal M}}

\newcommand\cL{{\cal L}}

\newcommand\cO{{\cal O}}

\newcommand\Tr{{\rm Tr}}

\begin{document}

\vspace*{1cm}

\begin{center}
{\bf \Large
On $SL(2,R)$ symmetry in nonlinear electrodynamics theories\\
% at order $O(\alpha'^2)$
}

\vspace*{1cm}

{{Komeil Babaei Velni}$^{{\,}{a,}}$\footnote{babaeivelni@guilan.ac.ir} and  H. Babaei-Aghbolagh$^{{\,}{b,}}$}\footnote{hossein.babaei66@gmail.com}\\
\vspace*{2cm}
{{}$^{{\,}{a}}\,$ Department of Physics, University of Guilan,\\ P.O. Box 41335-1914, Rasht, Iran}
\\
\vspace{.5cm}
{{}$^{{\,}{b}}\,$ Department of Physics, Ferdowsi University of Mashhad,\\ P.O. Box 1436, Mashhad, Iran, %P.O. Box 1436,
 Mashhad, Iran}
\\
\vspace{2cm}

\end{center}

\begin{abstract}
\baselineskip=18pt
Recently, it has been observed that the Noether-Gaillard-Zumino (NGZ) identity holds order by order in $\alpha'$ expansion in nonlinear electrodynamics theories as Born-Infeld (BI) and  Bossard-Nicolai (BN). The nonlinear electrodynamics theory that couples to an axion field is invariant under the $SL(2,R)$ duality  in all orders of $\alpha'$ expansion in the Einstein frame. In this paper we show that there are the  $SL(2,R)$ invariant forms of the energy momentum tensors of axion-nonlinear electrodynamics theories in the Einstein frame. These $SL(2,R)$ invariant structures appear in the energy momentum tensors of BI and BN theories at  all orders of $\alpha'$ expansion.  The $SL(2,R)$ symmetry appears in the BI and BN Lagrangians as a multiplication of Maxwell Lagrangian and a series of  $SL(2,R)$ invariant structures.

\end{abstract}
\vskip 0.5 cm

%Keywords:T-duality, S-matrix, Coupling

\vfill
\setcounter{page}{0}
\setcounter{footnote}{0}
\newpage
\section{Introduction} \label{intro}

Duality transformations in nonlinear electrodynamics has been studied in \cite{Aschieri:2008}. Classical electromagnetism is the most familiar duality-invariant theory. The Maxwell’s Hamiltonian and the equations of motion are invariant under rotations. Note that the Lagrangian, however, is not invariant. This brings that nonlinear deformations of Lagrangian will require modifications which are also non-invariant.

Commonly, duality transformations may be found out in the path integral as a Legendre transform. Given some Lagrangian $L(F)$ depending only on the field strength of a vector field, one can construct\cite{Carrascoa:1108}
\beqa
 {\tilde{\cL(F,G)}}=\cL(F)-\frac{1}{2}\epsilon^{abcd}F_{ab}\partial_{c}\tilde{A}_{d}
\eeqa
in which $F$ is treated as a fundamental field. The classical equations of motion for $F$ require that $G_{ab}=\partial_{a}{\tilde{A}}_{b}-\partial_{b}{\tilde{A}}_{a}$ is related to $F$ by
\beqa
G_{ab}=-2\frac{\partial \cL(F)}{\partial F^{ab}}\labell{G}
\eeqa
 through $G_{ab}=-\frac{1}{2}\epsilon_{abcd}{\tilde{G}}^{cd}$ and $\tilde{G}^{ab}=\frac{1}{2}\epsilon^{abcd} G_{cd}$. 

Consistency of the duality constraint can be expressed as a requirement in which the Lagrangian must transform under
duality in a particular way, defined by the Noether-Gaillard-Zumino (NGZ) identity \cite{Gaillard:1981}.% In the case of a U(1) duality, NGZ identity simplifies to the following relation:
%\beqa
% G\tilde{G}-\tilde{F}\tilde{F}=0\labell{ngz}
%\eeqa

We consider $\cL_{inv}$ that is invariant under the following transformation:
\beqa\labell{TE}
\delta \left(\begin{array}{c}
 \tilde{F} \\G \\
 \end{array}
 \right)=\left(\begin{array}{c}
 A\,\,\,\,\,\,\,B \\C\,\,\,\,\,\,\,\,D
 \end{array}
 \right)\left(\begin{array}{c}
 \tilde{F} \\G \\
 \end{array}
 \right)
 \eeqa
 
 where $A^T=-D$, $B^T=B$ and $C^T=C$ are the infinitesimal parameters of the transformations\cite{Carrascoa:1108}.
  By applying the duality symmetry, the NGZ identity could be written as following\cite{Carrasco:2012}:
\beqa
\cL=\cL_{inv}-\frac{1}{4} G_{ab}F^{ab}\labell{l}.
 \eeqa
When the theory only has linear duality (e.g. only $F^2$ terms in the action) $\delta{L_{inv}}/{\delta F}$ vanishes. So it could be found that any higher order dependence ($F^4, F^6,...$) must be part of $\cL_{inv}$. The NGZ identity, along with \reef{G} can be solved to find $G(F)$ and various Lagrangians which provide a duality symmetry between equations of motion and Bianchi identities. We will discuss two cases of nonlinear deformations of the Maxwell theory that depend only on $F'$s without derivatives.

The NGZ identity has a significant consequence for the duality rotations properties of the energy-momentum tensor. The energy-momentum tensor, which can be obtained as the variational derivative of the Lagrangian with respect to the gravitational field, is invariant under duality transformation\cite{Gaillard:9712}. It has been found as\cite{Gibbons:9509}
\beqa
T_{ab}=g_{ab}\cL+{{{G}_{a}}^{}}^{c}  F_{bc}\labell{Tab}
\eeqa
Considering this relation and \reef{l}, it could be found that $\cL_{inv}={1/4}  {{T}_{a}}^{{}{a}}$.

The Born-Infeld Lagrangian density can be written intelligently in term of the square root of a determinant\cite{Zwiebach:2004}:
\beqa
L_{BI}= \sqrt{-\det\left(\eta_{ab}+F_{ab}\right)}-1\labell{LBI}
\eeqa

where the fundamental $(scale)^2 (=T^{-1} = 2\pi \alpha'$ in the string theory context) has been set equal to 1, has many considerable aspects, including electro-magnetic duality symmetry \cite{Tseytlin:9908}.
Since in four dimensions:
\beqa
 -\det\left(\eta_{ab}+F_{ab}\right)=\frac{1}{2}F_{ab}F^{ab}-\frac{1}{16}(F_{ab}{\tilde{F}}^{ab})^{2}, \,\,\,\,\,\,\,\,\,\,\,\,\,\,\,\,\,\,\,\,   {\tilde{F}}^{ab}=\frac{1}{2}\epsilon^{abcd}F_{cd},
\eeqa
$L_{BI}$ interpolates between the Maxwell Lagrangian $\frac{1}{4}F_{ab}F^{ab}$ for small $F$ and the total derivative (topological density) $\frac{i}{4}(F_{ab}{\tilde{F}}^{ab})$ for large $F$. One can find an appropriate form in Euclidean signature as:
\beqa
&&L_{BI}=\sqrt{(1+F_2)^2+2F_4}-1=F_2+F_4[1+O(F^2)],\labell{F2}
\eeqa
where $F_2=\frac{1}{4}F_{ab}F^{ab}$ and $F_4=\frac{-1}{8}[F_{ab}F^{bc}F_{cd}F^{da}-\frac{1}{4}(F_{ab}F^{ab})^2]$. 

That $L_{BI} = F_2 +F_4 + O(F^6)$ is true in all dimensions, however it is only in $D=4$ that all higher order terms are proportional to $F_4$.% This fact is reflected in the structure of the
%supersymmetric generalization of the $D = 4$ BI action.

%The $D=4$ Born-Infeld action is obviously symmetric under $F\leftrightarrow {\tilde{F}}$ and is also covariant under the electric-magnetic (or vector $\leftrightarrow$ vector) duality, as can be concluded from the structure of the equations of motion \cite{Gibbons:9506} (see also \cite{Gaillard:9712,Brace:9905}).

Adding axion and dilaton fields to the Born-Infeld electrodynamics, the electric-magnetic duality invariance may be extended to SL(2,R) S-duality, relevant to
string theory, which insinuates a strong-weak coupling duality of such theories\cite{Rasheed:9702}. This action is not invariant under the
S-duality, however, its equations of motion and energy-momentum tensor are invariant under the S-duality as it was shown for the electric-magnetic duality\cite{hosein:1304}. The SL(2,R) S-duality transformation holds order by order in $\alpha'$ and is nonperturbative in the string loop expansion \cite{Becker:2007}.

 We will investigate the S-dual structure of energy momentum tensor of nonlinear electrodynamics theory that couples to dilaton and axion fields in the Einstein frame. In fact, we are going to study  the energy momentum tensor order by order in $\alpha'$. We want to show that the energy momentum tensor constructed of SL(2,R) structures at any order in $\alpha'$. We will find the form of energy momentum tensor that is manifestly S-dual. The action of the theory that appears as a part of energy momentum tensor, can be written in term of a production of Maxwell action and a set of S-dual structures.
\\
%ما نتایج حاصل از این ناوردای را در گرانش و معادله انشتین ( ٠١٠١٢٢٩ معادله ٢.۵ و SL 2) ;R)
%ناوردا برای کرویچر و تانسور ریچی در SL 2) ;R رفرنس ١٧ این کار)برسی میکنیم که منجر به یافتن ساختارھای (
%می شود. ھر مرتبه از
%Despite the highly non-linear nature of  the equations of motion have an exact
%SO(2)  [6].

As was noted in \cite{Gaillard:1981} that the invariance of the energy momentum tensor and thus of
the corresponding Lagransian should imply the invariance of the S-matrix. So one expects
to find the S-matrix elements in term of the $SL(2,R)$ invariant structures that appear in
the energy momentum tensors. On the other hand, it has been found that the tree-level S-matrix elements of some scattering amplitudes including the loops and the nonperturbative effects become symmetric under the $SL(2,Z)$ transformation. Using the Ward identity corresponding to the global S-duality transformations, this point can be extended to all S-matrix as well \cite{Garousi:1108}. It could be expected, considering the $SL(2,R)$ invariant stractures of the energy momentum tensors and the Ward identity corresponding to the global S-duality transformations, have been used as guiding principles to find the S-matrix elements\cite{Hosein:1600}. The action of linearized $SL(2,R)$ invariance on S-matrix of scalars and vectors for $D_3$-brane in tree-level open/closed string theory was studied in \cite{hosein:1304, Garousi:1108}.
  
% and it was found that there are in fact an infinite
%number of such duality invariant theories. However, Born-Infeld, along
%with Maxwell theory, is the only one which is known in an exact closed form.
The theory that comes from nontrivial nonlinear deformation of classical electrodynamics, which is consistent with NGZ identity, is Bossard-Nicolai theory. This theory is produced in which the NGZ condition work order by order in $\alpha'$. This theory is equal to Born-Infeld theory up to $O(F^6)$. Indeed, the Bossard-Nicolai theory differs from  BI theory starting at $O(F^8)$\cite{Carrascoa:1108}. It will be expected that the action and the energy momentum tensor of BN theory appear in term of SL(2,R) structure. In fact, we extend our results from the BI theory to the BN theory. 

The outline of the paper is as follows: We begin in section 2 by reviewing the SL(2,R) transformations of bosonic fields. We investigate the behavior of BI energy momentum tensor under the SL(2,R) transformations. In section 3, we extend our calculation in previous section to all orders of $\alpha'$. We use these considerations to find the BN theory in term of relevant SL(2,R) invariant structure.

\section{SL(2,R) invariant structure}

In this section we are going to construct relevant structures which are manifestly invariant under SL(2,R) transformation. The nonlinear transformation of the gauge field and the axion-dilaton, $\tau=C_0+ie^{-\phi_0}$
are given by\cite{Gibbons:9509}
\begin{eqnarray}
F_{ab}&\rightarrow &sF_{ab}+r\tilde{G}_{ab} \nonumber\\
G_{ab}&\rightarrow &pG_{ab}-q\tilde{F}_{ab}\,\,\,\,\,\,;\,\,\,\,\,\,\,\,\,\tau\rightarrow \frac{p\tau+q}{r\tau+s}\labell{axidil}
\end{eqnarray}

where the antisymmetric tensor $G_{ab}$ is the one appears in \reef{G}. It could be written the transformation of gauge field
as
\begin{eqnarray}
\cF_{a b}\equiv\left(
\begin{array}{c}
\tilde{F}_{ab} \cr  \
 G_{a b}
\end{array}
\right) \rightarrow  (\Lambda^{-1})^T\left(
\begin{array}{c}
\tilde{F}_{ab} \cr  \
G_{a b}
\end{array}
\right)\,\,\,\,\,\,\,;\,\,\,\,\,\,\,\,\,\Lambda=\left(
\begin{array}{cc}
p&q  \\
r&s
\end{array}
\right)\in\, {SL(2,R)}\labell{sl2r}
\end{eqnarray}

The S-duality transformation on the background fields $\phi_0$ and $C$ is
\beqa
{\cal M}\rightarrow \Lambda {\cal M}\Lambda ^T \,\,\,\,\,\,\,;\,\,\,\,\,\,\,\,\,  {\cal M}=e^{\phi_0}\pmatrix{|\tau|^2&C_0 \cr 
C_0&1}\labell{M}\labell{MM}
\eeqa

Using the above transformations, one can find that the structure $\cF^T\cM\cF $ is a $SL(2,R)$ invariant structure.

It could be useful to consider the Lagrangian at the presence of axion and dilaton couplings in the following form which the contribution of axion coupling is separated.
\beqa
\cL=\cL^{'}+C_0z
\eeqa 
where the Lagrangian was constructed of the following two possible Lorentz invariants ($\cL (t , z)$):
\beqa
t=\frac{1}{4}F^{a b} F_{a b} \, \, \, \, , \, \, \, z=\frac{1}{4}F^{a b} \tilde{F}_{ab}\labell{tz}
\eeqa
The antisymmetric tensor $G_{ab}$ then seperates as $G_{ab}=G^{'}_{ab}-C_0\tilde{F}_{ab}$ where $G^{'}_{ab}$ could be an arbitrary nonlinear function of $F$ which is $G^{'}_{ab}=-2\frac{\prt \mathcal{L}^{'}}{\prt F^{ab}}$. 

By this consideration one gets the $SL(2,R)$ invariant structure in the following form:
\beqa
(\cF^T)_a{}^c\cM_0\cF_{bc}=e^{-\phi_0}\,\tilde{F}_a{}^c\,\tilde{F}_{bc}+ e^{\phi_0}\, G^{'}_{a}{}^c\, G^{'}_{bc}
\eeqa 
It is easy to check that the above structure is invariant under the linear transformations $G^{'}   \longrightarrow  \tilde{F}$ , $\tilde{F} \longrightarrow   -G^{'}$ and $e^{-\phi_0}  \longrightarrow  e^{\phi_0}$.
%\beqa
%G^{'}   \longrightarrow  \tilde{F} \,\,\,\,\, ;\,\,\,\,\, \tilde{F} \longrightarrow   -G^{'}  \,\,\,\,\,  ;        \,\,\,\,\, e^{-\phi_0}  \longrightarrow  e^{\phi_0} \nonumber
%\eeqa 

\section{Nonlinear electrodynamics theories}

In this section we review the nonlinear electodynamics theories that couple to axion field and investigate their behaviour under the $SL(2,R)$ symmetry. The actions of these nonlinear theories are the generalized form of Maxwell action. In fact, it was demonstrated that the Maxwell action is the leading-order
term in the expansion in $F$ of nonlinear electodynamics theories. These actions satisfy the NGZ identity, however they are not invariant under the S-duality transformation. In following, we consider the S-duality transformations in the axion-Maxwell theory and show the energy momentum tensor of this theory is invariant under the $SL(2,R)$ symmetry. We extend this consideration to nonlinear theories: Axion-Born-Infeld and Axion-Bossard-Nicolai.

\subsection{Axion-Maxwell theory}
The simplest example of duality invariant theories is Maxwell's electromagnetism. The Maxwell action can be generalized to the case of the presence of background axion field\footnote{We have also introduced a background dilaton field $\phi_0$ with $e^{\phi_0/2}$ playing the role of an effective gauge coupling constant.}\cite{Gibbons:9509} 
\beqa
\cL=-e^{-\phi_0}t+C_0z
\eeqa
where $t$ and $z$ are the ones appear in \reef{tz}. According to \reef{Tab} and \reef{G}, the energy momentum tensor of the above action could be found in the following form:
\beqa
T_{ab}= e^{- \phi_0 }\left[- \frac{1}{4}F_{cd}F^{cd}g_{ab}+F_{a}{}^cF_{bc}\right]\labell{Tm}
\eeqa
On the other hand, replacing in the S-duality transformation \reef{sl2r} and \reef{MM}  the relevant $G_{ab}$  from \reef{G}, one finds the following $SL(2,R)$ invariant structure:
\beqa
(\cF^T)_a{}^c\cM_0\cF_{bc}=e^{-\phi_0}[\tilde{F}_a{}^c\tilde{F}_{bc}+F_{a}{}^cF_{bc}]=e^{-\phi_0}[-\frac{1}{2}F_{cd}F^{cd}g_{ab}+2F_{a}{}^cF_{bc}]\labell{sl2rm}
\eeqa
where we use of the identity $\tilde{\tilde{F}}_{ab}=-F_{ab}$ and 
\beqa
\eps^{abcd}\eps^{efgh}=-\left|\begin{array}{cccc}
\eta^{ae}& \eta^{af}& \eta^{ag}&\eta^{ah}\cr  \
 \eta^{be}& \eta^{bf}& \eta^{bg}&\eta^{bh}\cr \
 \eta^{ce}& \eta^{cf}& \eta^{cg}&\eta^{ch}\cr \
\eta^{de}& \eta^{df}& \eta^{dg}&\eta^{dh}
\end{array}\right|.\labell{zero2n} 
\eeqa

Comparing the energy momentum tensor \reef{Tm} and the structure \reef{sl2rm}, the energy momentum tensor in the form that is manifestly $SL(2,R)$ invariant could be presented as:
\beqa
{T_{ab}}&=&\frac{1}{2} (\cF^T)_a{}^c\cM_0\cF_{bc}\labell{tmaxwell}
%&=&\frac{1}{2}\bigg(e^{-\phi_0}\tilde{F}_a{}^c\tilde{F}_{cb}+ e^{\phi_0} G^{'}_{a}{}^c G^{'}_{cb}\bigg)\labell{tmaxwell}
\eeqa
%One can check that the energy momentum tensor is invariant under
 It is obvious that $Tr(\cF^T\cM_0\cF)$ will thus be zero (In the next sections, we see that the trace of $SL(2,R)$ invariant structure is nonzero for nonlinear electrodynamics theories). 
%The non-linear electrodynamic (as Born-Infeld theory) by including a coupling to a scalar dilaton field $\phi_0$ and a pseudo-scalar axion field $a$ has been studied in \cite{Gibbons:9506,Gibbons:9509}. The resulting theory admited an SL(2,R) electric-magnetic duality which mixes the electromagnetic field equations with the Bianchi identities and also transforms the axion and dilaton. 
We will find the energy momentum tensors corresponding to the BI and BN theories in the forms which are manifestly $SL(2,R)$ invariant.

\subsection{Axion-Born-Infeld theory}

The Born-Infeld theory, perhaps is the simplest nonlinear deformation of Maxwell’s theory. The Axion-Born-Infeld Lagrangian, the general nonlinear electromagnetic theory couple to the axion field that the equations of motion are $SL(2,R)$ invariant, in the Einstein fram is:
\beqa
\cL ={\cL}_{BI}+C_0 z={g}^{-2} \bigg[1-\sqrt{1+2{g}^{2}e^{- \phi_0 }t -{g}^{4}e^{- 2\phi_0 }z^2}\bigg]+C_0 z\labell{LL}
\eeqa
where $g=2\pi \alpha^{'}$. Regardless the axion term, it is clear that classical electromagnetism corresponds to $g^2\rightarrow 0$. In the following calculation, we set $\alpha'=1/{2\pi}$.
Using \reef{G}, we find the following expression for G:
\beqa
G_{ab}=\frac{e^{- \phi_0 }F_{ab}-e^{-2 \phi_0}z\tilde{F}_{ab}}{\sqrt{1+2e^{- \phi_0}t -e^{-2\phi_0}z^2}}-C_0\tilde{F}_{ab}\labell{Gam}
\eeqa

One can expand the Born-Infeld Lagrangian as:
\beqa
\mathcal{L}_{BI}=-t e^{-\phi_0 }+\frac{1}{2} e^{-2 \phi_0 } \left(t^2+z^2\right)-\frac{1}{2} t e^{-3 \phi_0 } \left(t^2+z^2\right)+\frac{1}{8} e^{-4 \phi_0 } \left(t^2+z^2\right) \left(5 t^2+z^2\right)+\ldots
\eeqa

Using \reef{zero2n}, we can write the field variable $z^2$ in term of the following trace terms:
\beqa
z^2=\frac{1}{4} Tr(F\inn F\inn F\inn F)-\frac{1}{8} Tr(F\inn F)^2\labell{z2}
\eeqa 

To simplify, we choose the $SL(2,R)$ matrix $\Lambda$ in \reef{sl2r} which has been made of the components $p=0, q=1, r=-1$ and $s=0$. This consideration brings the following linear S-duality transformations
%\beqa
%e^{- \phi_0} F_{ab}\to \tilde{F}_{ab}\,\,\,\,\,\,\,\,\,;\,\,\,\,\,\,\,\,\,F_{ab}\to e^{-\phi _0}\tilde{F}_{ab}\,\,\,\,\,\,\,\,\,;\,\,\,\,\,\,\,\,\, \phi_0\rightarrow -\phi_0\labell{lsd}
%\eeqa
\beqa
   F_{ab}\to e^{- \phi _0} \tilde{F}_{ab} \,\,\,\,\,   \,\,\,\,\,   \,\,\,\,\,   ; \,\,\,\,\,\,\,\,\tilde{F}_{ab}\to -e^{- \phi _0} F_{ab}  \,\,\,\,\,   \,\,\,\,\,   \,\,\,\,\,   ; \,\,\,\,\,\,\,\, \phi_0\rightarrow - \phi_0
\eeqa
According to \reef{zero2n} and the following trace S-duality transformations
\beqa
e^{-\phi_0}Tr(F\inn F)\to -e^{-\phi_0}Tr(F\inn F)\,\,\,\,\,\,\,\,\,;\,\,\,\,\,\,\,\,\, e^{-2 \phi_0}Tr(F\inn F\inn F\inn F)\rightarrow e^{-2 \phi_0} Tr(F\inn F\inn F\inn F) 
\eeqa
it will be revealed that $e^{-2 \phi _0} z^2$ is invariant under the linear S-duality transformation. Replacing the field variable $z^2$ from \reef{z2} in the Born-Infeld Lagrangian, one finds the expansion of $\cL_{BI}$ at the all levels of $F^{2n}$ in term of trace structures $Tr(F\inn F)$ and $Tr(F\inn F\inn F\inn F)$.
\beqa
{\cL}_{BI}&=& e^{- \phi_0 }\left[ \frac{1}{4}Tr(F\inn F)\right] \nonumber\\
&+&e^{-2 \phi_0 }\left[\frac{1}{8}\Tr(F\inn F\inn F\inn F)-\frac{1}{32}\Tr(F\inn F)^2\right]\nonumber\\
 &+&e^{-3 \phi_0 } \bigg[-\frac{1}{32}\Tr(F\inn F\inn F\inn F)\Tr(F\inn F)+\frac{1}{128}\Tr(F\inn F)^3 \bigg] \nonumber\\
 &+&e^{-4 \phi_0 } \bigg[\frac{1}{128}\Tr(F\inn F\inn F\inn F)^2+\frac{1}{256}\Tr(F\inn F)^2\Tr(F\inn F\inn F\inn F)-\frac{3}{2048}\Tr(F\inn F)^4\bigg]  \nonumber\\
&+&\cO (F^{10})+\ldots
\eeqa
It is easy to show that the above Lagrangian $\cL_{BI}$ in the all levels of $F^{2n}$ is self-dual and anti-self-dual under the linear S-duality transformation when n is even and odd respectively.  
%\newpage
By expanding the denominator in \reef{Gam}, we can find the $G_{ab}$ in the  following expression
\beqa
G_{ab}&=&G^{'}_{ab}-C_0\tilde{F}_{ab}\nonumber\\
&=&e^{- \phi_0 } F_{ab}-C_0\tilde{F}_{ab}+e^{- 2\phi_0 }\bigg( F_a\inn F\inn F_b-\frac{1}{4}Tr(F\inn F)F_{ab}\bigg)\nonumber\\
&+&e^{- 3\phi_0 }\bigg(\frac{1}{8}Tr(F\inn F\inn F\inn F)F_{ab}+\frac{1}{4}Tr(F\inn F)F_a\inn F\inn F_b-\frac{3}{32} {Tr(F\inn F)}^2F_{ab}\bigg)\nonumber\\
&+&e^{-4 \phi_0 } \bigg(\frac{1}{8}Tr( F\inn F\inn F\inn F)F_a\inn F\inn F_b+\frac{1}{32}Tr(F\inn F) Tr(F\inn F\inn F\inn F)F_{ab}\nonumber\\
&+&\frac{1}{32}Tr(F\inn F)^2F_a\inn F\inn F_b-\frac{3}{128} Tr(F\inn F)^3F_{ab}\bigg)+\cO (F^{9})+\ldots\labell{GGabi}
\eeqa 
Replacing the above expansions of $\cL_{BI}$ and $G_{ab}$ in \reef{Tab}, the energy momentum tensor becomes 
\beqa
T_{ab}&=&
 e^{- \phi_0 }\left[ \frac{1}{4}g_{ab}Tr(F\inn F)-F_a\inn F_b\right]\nonumber\\
 &+&e^{-2 \phi_0 }\left[g_{ab}\left(\frac{1}{8}Tr(F\inn F\inn F\inn F)-\frac{1}{32}Tr(F\inn F)^2\right)-F_a\inn F\inn F\inn F_b+ \frac{1}{4} Tr(F\inn F)F_a\inn F_b\right]\nonumber\\
 &+&e^{-3 \phi_0 } \bigg[g_{ab}\left(\frac{1}{32}Tr(F\inn F\inn F\inn F)Tr(F\inn F)-\frac{1}{128}Tr(F\inn F)^3\right)\nonumber\\
&-&\frac{1}{8}Tr(F\inn F\inn F\inn F)F_a\inn F_b- \frac{1}{4} Tr(F\inn F)F_a\inn F\inn F\inn F_b+\frac{3}{32}Tr(F\inn F)^2F_a\inn F_b \bigg]\nonumber \\
&+&e^{-4 \phi_0 }\bigg[g_{ab} \left(\frac{1}{128}Tr(F\inn F\inn F\inn F)^2+\frac{1}{256}Tr(F\inn F)^2Tr(F\inn F\inn F\inn F)-\frac{3}{2048}Tr(F\inn F)^4\right)\nonumber\\
&-&\frac{1}{8}  Tr(F\inn F\inn F\inn F)F_a\inn F\inn F\inn F_b-\frac{1}{32}  Tr(F\inn F) Tr(F\inn F\inn F\inn F)F_a\inn F_b\nonumber\\
&-&\frac{1}{32} Tr(F\inn F)^2F_a\inn F\inn F\inn F_b +\frac{3}{128}  Tr(F\inn F)^3F_a\inn F_b\bigg]+\cO (F^{10})+\ldots\labell{Tborn}
\eeqa
In fact, the above BI energy momentum tensor is the generalization of Maxwell energy momentum tensor \reef{Tm}. It is not clear how the above tensor treat under the S-dual transformation. To answer this question, let us try to present the above BI energy momentum tensor in the form that is $SL(2,R)$ invariant. So we calculate appropriate $SL(2,R)$ invariant structure and should be able to write the energy momentum tensor in term of this structure. To this end, we make the nonlinear  $SL(2,R)$ invariant structure by replacing in \reef{sl2r} the antisymmetric tensor $G_{ab}$ from \reef{GGabi}. This structure becomes:
\beqa
(\cF^T)_a{}^c\cM_0\cF_{cb}&=&e^{-\phi_0 }\bigg[-\frac{1}{2}  g_{ab} Tr(F\inn F)+2  F_a\inn F_b\bigg] \nonumber\\
&+&e^{- 2\phi_0}\bigg[-\frac{1}{2}   Tr(F\inn F)F_a\inn F_b+2  F_a\inn F\inn F\inn F_b\bigg]\nonumber\\
&+&e^{-3 \phi_0 }\bigg[ F_a\inn F\inn F\inn F\inn F\inn F_b +\frac{1}{4}  Tr(F\inn F\inn F\inn F)F_a\inn F_b -\frac{1}{8} Tr(F\inn F)^2F_a\inn F_b\bigg]\nonumber\\
&+&e^{-4 \phi_0 }\bigg[\frac{1}{2} Tr(F\inn F) F_a\inn F\inn F\inn F\inn F\inn F_b -\frac{1}{4}   Tr(F\inn F)^2F_a\inn F\inn F\inn F_b\nonumber\\
&+&\frac{1}{2}   Tr(F\inn F\inn F\inn F)F_a\inn F\inn F\inn F_b\bigg]+\cO (F^{10})+\ldots\labell{bisl2r}
\eeqa
To build the energy momentum tensor \reef{Tborn} in term of the above $SL(2,R)$ invariant structure we need to apply valuable identities at the level of $n\geq 3$. These identities for $n=3$ are as following:
\beqa
Tr(F\inn F\inn F\inn F\inn F\inn F) -\frac{3}{4} Tr(F\inn F\inn F\inn F)Tr(F\inn F) +\frac{1}{8}   {Tr(F\inn F)}^3&=&0\nonumber\\
F_a\inn F\inn F\inn F\inn F\inn F_b-\frac{1}{2}   Tr(F\inn F)F_a\inn F\inn F\inn F_b+\frac{1}{4}F_a\inn F_b[\frac{1}{2} {Tr(F\inn F)}^2- Tr(F\inn F\inn F\inn F)]&=&0\nonumber\nonumber\\\labell{identity}
%\eeqa
%\beqa
\eeqa

Using these identities one can write $T_{ab}$ in the form that is manifestly $SL(2,R)$ invariant.
\beqa 
T_{ab}=g_{ab}-\bigg[ g_{ab}+\frac{1}{2}(\cF^T)_a{}^c\cM_0\cF_{cb}\bigg]\sum _{m=0}^{\infty }a_m \,\,(-1)^m \,\,2^{-4m} \,\, {Tr(\cF^T\cM_0\cF)}^m\labell{tbi}
\eeqa
where
\beqa
a_0=1\,\,\,\,;a_1=1\,\,\,\,;a_2=\frac{3}{2}\,\,\,\,;a_3=\frac{5}{2}\,\,\,\,;a_4=\frac{35}{8}\,\,\,\,;a_5=\frac{63}{8}\,\,\,\,;\ldots\nonumber
\eeqa
 One can find the above energy momentum tensor which is expanded in term of the level of coupling factor $e^{-n\phi_0}$, where $n=2m+1$. We calculate the expansion factors up to $m=5$ or up to level of $e^{-11\phi_0}$ $ (\cO(F^{22}))$ . It is clear that the Born-Infeld energy momentum tensor leads to  the Maxwell energy momentum tensor \reef{tmaxwell} for $m=0$.
\small{\small{
%\beqa
%T_{ab}=g_{ab}-\bigg[ g_{ab}+\frac{1}{2}\bigg(e^{-\phi_0}\tilde{F}_a{}^c\tilde{F}_{cb}+ e^{\phi_0} G^{'}_{a}{}^c G^{'}_{cb}\bigg)\bigg]\sum _{m=0}^{\infty }a_{m} (-1)^m 2^{-4m} {Tr\bigg(e^{-\phi_0}\tilde{F}\tilde{F}+ e^{\phi_0} G^{'} G^{'}\bigg)}^m
%\eeqa}}

One may find the manifestly $SL(2,R)$ invariant form of the energy momentum tensor in terms of the different $SL(2,R)$ invariant structures  as $Tr({\cal F}^T{\cal M}_0{\cal F}{\cal F}^T{\cal M}_0{\cal F})$, $Tr({\cal F}^T{\cal M}_0{\cal F}{\cal F}^T{\cal M}_0{\cal F}{\cal F}^T{\cal M}_0{\cal F})$ and so on. It could be shown that these structures can be written in terms of powers of  $Tr({\cal F}^T{\cal M}_0{\cal F})$ (see appendix). From these considerations, the manifestly $SL(2,R)$ invariant form of the energy momentum tensors in this paper constructed out of the structure $Tr({\cal F}^T{\cal M}_0{\cal F})$ would be unique.

At the end of this section, we are going to find the representation of Born-Infeld Lagrangian that deduces from the relevant $SL(2,R)$ invariant energy momentum tensor \reef{tbi}. According to \reef{Tab} (and the energy-momentum tensor, which can be obtained as the variational derivative of the Lagrangian with respect to the gravitational field), one can find that only the first term in \reef{bisl2r} contributes to the $SL(2,R)$ invariant structure which appears in \reef{tbi}. Considering this, one can find the following representation for Born-Infeld Lagrangian:
 \beqa
 {\cal{L}}_{BI}=1-\big( 1+e^{- \phi_0 }t\big)\sum _{m=0}^{\infty }a_m \,\,(-1)^m \,\,2^{-4m}\,\, {Tr(\cF^T\cM_0\cF)}^m\labell{A}
 \eeqa
In this form of $\cL_{BI}$, it is obvious that the Lagrangian is constructed from two separable parts: invariant and non-invariant $SL(2,R)$ structures. As expected,  the $\cL_{BI}$ leads to Maxwell Lagrangian for $m=0$.

\subsection{Graviton-Axion-Born-Infeld theory}
It has been shown that  the Born-Infeld Lagrangian coupled to gravity $L(F_{ab},g_{cd})$  that the equations of motion, including the Einstein equations, are invariant under the  $SL(2,R)$ duality, has the following form\cite{Gibbons:9509}:
\beqa
\mathcal{L} =R+1-\sqrt{1+2e^{- \phi_0 }t -e^{- 2\phi_0 }z^2}+C_0 z\nonumber
\eeqa
In this section, we want to write the Einstein equation $R_{ab}-\frac{1}{2}Rg_{ab}= T_{ab}$ in the form that is manifestly $SL(2,R)$ invariant\footnote{we consider  the Einstein constant $G=1/{8\pi}$}. Since the metric is invariant under the  $SL(2,R)$ duality in the Einstein frame, one can expect that the scalar curvature and ricci tensor can be constructed in term of  $SL(2,R)$ invariant structures. Take the trace of the Einstein equation leads to $R=-{T_a}^a$. By considering \reef{tbi}, one can find  the scalar curvature in term of  $SL(2,R)$ invariant structure $Tr(\cF^T\cM_0\cF)$ as following:
\beqa
R&=&\frac{1}{4} \sum _{m=0}^{\infty } b_m (-1)^m 2^{-4 m}  {Tr(\cF^T\cM_0\cF)}^{m+1}\\
&& b_0= 1\,\,\,\,;b_1= \frac{1}{2}\,\,\,\,;b_2= \frac{1}{2}\,\,\,\,;b_3= \frac{5}{8}\,\,\,\,;b_4= \frac{7}{8}\,\,\,\,;b_5= \frac{63}{4}\ldots\nonumber
\eeqa
As we expected for the Maxwell theory, $m=0$, we get $R=0$. Replacing in the Einstein equation the above expansion and the expansion \reef{tbi}, the ricci tensor then becomes in term of following $SL(2,R)$ invariant structures:
\beqa
R_{ab}+g_{ab}&=&\left(g_{ab}+g_{ab}\frac{Tr(\cF^T\cM_0\cF) }{4}-\frac{(\cF^T)_a{}^c\cM_0\cF_{cb}}{2}\right)\sum _{m=0}^{\infty }a_m \,(-1)^m 2^{-4m}\, {Tr(\cF^T\cM_0\cF)}^m\nonumber
\eeqa
One can use the above considerations to write Riemann tensor  $R_{abcd}$ and any scalar structure as  $R_{ab}R^{ab}$ or $R_{abcd}R^{abcd}$ in term of $SL(2,R)$ invariant structures.

\subsection{Axion-Bossard-Nicolai theory}
Another NGZ-consistent theory is Bossard-Nicolai theory that comes from a distinct nontrivial nonlinear deformation of classical electrodynamics.
 We are going to show that the Lagrangian and the energy momentum tensor of BN theory appear in term of SL(2,R) structures.
 
 Using the definition of $G$ and $\tilde{G}$ according to \reef{G} and field variables \reef{tz}, the NGZ consistency could be presented as a differential equation\cite{Carrascoa:1108, Bossard:1105}.
 %\beqa
% {(({\prt}_{t}\mathcal{L})}^2-{({\prt}_{z}\mathcal{L})}^2-1)z-(2({\prt}_{z}\mathcal{L})({\prt}_{t}\mathcal{L}))t&=&0
% \eeqa
 Both Lagrangians $\cL_{BI}$ and $\cL_{BN}$ are the solution of NGZ differential equation. However unlike the BI Lagrangian, it is not known if the BN Lagrangian has a closed-form expression. Note that this Lagrangian differs from $\cL_{BI}$ starting at $\cO(F^6)$. In \cite{Carrascoa:1108}, this Lagrangian was calculated up to level of coupling factor $e^{-8 \phi_0 }$.
 \beqa
 {\mathcal{L}}_{BN}&=&-t e^{-\phi_0 }+\frac{1}{2} e^{- 2\phi_0} \left(t^2+z^2\right)-\frac{1}{2} t e^{-3 \phi_0 } \left(t^2+z^2\right)+\frac{1}{4} e^{-4 \phi_0 } \left(t^2+z^2\right) \left(3 t^2+z^2\right)\nonumber\\
&&-\frac{1}{8} t e^{-5 \phi_0 } \left(t^2+z^2\right) \left(11 t^2+7 z^2\right)+\frac{1}{32} e^{-6 \phi_0 } \left(t^2+z^2\right) \left(91 t^4+86 t^2 z^2+11 z^4\right)\nonumber\\
&&-\frac{1}{8} t e^{-7 \phi_0 } \left(t^2+z^2\right) \left(51 t^4+64 t^2 z^2+17 z^4\right)\nonumber\\
&&+\frac{1}{64} e^{-8 \phi_0 } \left(t^2+z^2\right) \left(969 t^6+1517 t^4 z^2+623 t^2 z^4+43 z^6\right)+\ldots
 \eeqa
 Now using \reef{z2}, one can expand the above Lagrangian at all orders of $\alpha^{'}$ in term of $Tr(F\inn F)$ and $Tr(F\inn F\inn F\inn F)$. Similar Axion-Born-Infeld Lagrangian \reef{LL}, the Axion-Bossard-Nicolai Lagrangian is $\mathcal{L}=\mathcal{L}_{BN}+C_0 z$.  
  Like the antisymmetric tensor \reef{GGabi}, we calculate the BN antisymmetric tensor $G_{ab}$ as following:
 \beqa
G_{ab}&=&G^{'}_{ab}-C_0\tilde{F}_{ab}\nonumber\\
&=&e^{- \phi_0 } F_{ab}-C_0\tilde{F}_{ab}+e^{- 2\phi_0 }\bigg( F_a\inn F\inn F_b-\frac{1}{4}Tr(F\inn F)F_{ab}\bigg)\nonumber\\
&+&e^{- 3\phi_0 }\bigg(\frac{1}{8}Tr(F\inn F\inn F\inn F)F_{ab}+\frac{1}{4}Tr(F\inn F)F_a\inn F\inn F_b-\frac{3}{32} {Tr(F\inn F)}^2F_{ab}\bigg)\nonumber\\
&+&e^{-4 \phi_0 } \bigg(\frac{1}{4}Tr(F\inn F\inn F\inn F) F_a\inn F\inn F_{b} -\frac{1}{64}Tr(F\inn F)^{3} F_{ab} \bigg) \nonumber\\
&+&e^{-5 \phi_0 } \bigg(\frac{15}{2048}Tr(F\inn F)^{4} F_{ab} -\frac{5}{128}Tr(F\inn F)^{3} F_a\inn F\inn F_{b} -\frac{15}{256}Tr(F\inn F\inn F\inn F) Tr(F\inn F)^{2} F_{ab} \nonumber\\
&+& \frac{7}{32}Tr(F\inn F\inn F\inn F) Tr(F\inn F) F_a\inn F\inn F_{b} +\frac{7}{128}Tr(F\inn F\inn F\inn F)^{2 } F_{ab} \bigg)\nonumber\\
&+& e^{-6 \phi_0 } \bigg(\frac{111  }{16384}Tr(F\inn F)^{5} F_{ab}-\frac{79  }{4096}Tr(F\inn F)^{4} F_a\inn F\inn F_{b}\nonumber\\
&-&\frac{79 }{2048}Tr(F\inn F\inn F\inn F) Tr(F\inn F)^{3} F_{ab} +\frac{31}{512}Tr(F\inn F\inn F\inn F) Tr(F\inn F)^{2} F_a\inn F\inn F_{b}\nonumber\\
&+&  \frac{31  }{1024}Tr(F\inn F\inn F\inn F)^2 Tr(F\inn F)F_{ab}+\frac{33}{256}Tr(F\inn F\inn F\inn F)^{2} F_a\inn F\inn F_{b} \bigg) \nonumber\\
&+& e^{-7 \phi_0 } \bigg(\frac{63  }{32768}Tr(F\inn F)^{6} F_{ab}-\frac{5  }{4096}Tr(F\inn F)^{5} F_a\inn F\inn F_{b}-\frac{25}{8192}Tr(F\inn F\inn F\inn F) Tr(F\inn F)^{4} F_{ab} \nonumber\\
&-&\frac{21}{512}Tr(F\inn F\inn F\inn F) Tr(F\inn F)^{3} F_a\inn F\inn F_{b} -\frac{63 }{2048}Tr(F\inn F\inn F\inn F)^{2} Tr(F\inn F)^{2}F_{ab} \nonumber\\
&+&\frac{51}{256}Tr(F\inn F\inn F\inn F)^{2} Tr(F\inn F) F_a\inn F\inn F_{b} +\frac{17}{512}Tr(F\inn F\inn F\inn F)^{3} F_{ab} \bigg)\nonumber\\
&+&e^{-8 \phi_0 } \bigg(-\frac{83  }{131072}Tr(F\inn F)^{7}F_{ab}+\frac{271  }{65536}Tr(F\inn F)^{6}F_a\inn F\inn F_{b}\nonumber\\
&+&\frac{813  }{65536}Tr(F\inn F\inn F\inn F) Tr(F\inn F)^{5}F_{ab}-\frac{103}{2048}Tr(F\inn F\inn F\inn F) Tr(F\inn F)^{4}F_a\inn F\inn F_{b}\nonumber\\
&-&\frac{103  }{2048}Tr(F\inn F\inn F\inn F)^{2} Tr(F\inn F)^{3}F_{ab}+\frac{483 }{4096}Tr(F\inn F\inn F\inn F)^{2} Tr(F\inn F)^{2} F_a\inn F\inn F_{b}\nonumber\\
&+&\frac{161 }{4096}Tr(F\inn F\inn F\inn F)^{3} Tr(F\inn F) F_{ab}+\frac{43}{512}Tr(F\inn F\inn F\inn F)^{3} F_a\inn F\inn F_{b} \bigg)+\cO (F^{17})+\ldots\nonumber
 \eeqa
 
 One can construct the $SL(2,R)$ invariant structures for this theory by replacing in \reef{sl2r} the above antisymmetric tensor. Using \reef{Tab}, the BN energy momentum tensor could be derived. 
  The form of the energy momentum tensor in term of the invariant structure then becomes:
 \beqa
 T_{ab}&=&-\frac{1}{4}(\cF^T)_a{}^c\cM_0\cF_{cb}\nonumber\\
 &&-\frac{1}{4}\bigg[(\cF^T)_a{}^c\cM_0\cF_{cb}-\frac{Tr(\cF^T\cM_0\cF) g _{ab}}{4}\bigg]\sum _{m=0}^{\infty } (-1)^m 2^{-4 m} a_m Tr(\cF^T\cM_0\cF)^m\nonumber\\
&&a_0=1\,\,\,\,;a_1=2,\,\,\,;a_2=1\,\,\,\,;a_3=\frac{1}{2};\ldots
 \eeqa
where we calculate the coefficients of series up to coupling factor $e^{-8 \phi_0 }$. 
%Like the form of $L_{BI}$ \reef{A},
Using \reef{Tab}, we find the BN Lagrangian in term of the $SL(2,R)$ invariant structure as following:
\beqa
{\mathcal{L}}_{BN}=-\frac{1}{2} e^{-\phi_0 } t-\frac{1}{16} \bigg[8 e^{-\phi_0 } t-Tr(\cF^T\cM_0\cF)\bigg]\sum _{m=0}^{\infty } (-1)^m 2^{-4 m} a_m Tr(\cF^T\cM_0\cF)^m
\eeqa

This ends our illustration of the behavior of nonlinear electrodynamics theories at the presence of dilaton and axion fields under the $SL(2,R)$ symmetry. We could write the actions of these theories in term of $SL(2,R)$ invariant structure. We could explicitly check the invariance of the energy momentum tensor of Axion-
dilaton-nonlinear electrodynamics theories under the S-duality transformation and found relevant $SL(2,R)$  invariant structures.

%Investigating the energy momentum tensor of these theories, we could explicitly check
%its invariance and in fact answer the question that proposed in \cite{Gaillard:9712}: How could be manifested the invariance of the energy momentum tensor of Axion-dilaton-nonlinear electrodynamics theories under the S-duality transformation?
The nonlinear electrodynamics theory has been expressed as a leading term in the low-energy effective action of string theory. The BI action may be defined as the action representing the $D_3$-branes in 4 dimensions\cite{Tseytlin:9908}. The equations of motion and the energy momentum tensor of the effective action of $D
_3$-branes are invariant under $SL(2,R)$ transformation\cite{Green:1996}. It is expected that this symmetry will be held at higher order fields. The effective action of $D_3$-brane has been presented in \cite{Garousi:1511} in term of two linear S-dual invariant $Q$-tensors in which $Q=\partial F\partial F+\partial \tilde{F}\partial \tilde{F}$. On the other hand, we have shown in this paper that the $SL(2,R)$ invariant structures appear nonlinearly in the effective action of $D_3$-brane. So using nonlinear invariant structure as $Q'=\partial G'\partial G'+\partial \tilde{F}\partial \tilde{F}$
% in the effective action of $D_3$-brane
, it will be expected that one can derive more higher order couplings.  

In the approach suggested in \cite{Ivanov:2001} just mentioned the whole information about the given duality invariant system is encoded, in a closed form, in some invariant function of tensorial auxiliary fields, while the representation of the relevant Lagrangians as infinite series over powers of the Maxwell field strength arises as a result of elimination of these auxiliary fields by their algebraic equation of motion. It would be interesting to find a closed form of the relevant energy momentum tensor in terms of their $SL(2,R)$ invariants using this formulation.

{\bf Acknowledgments}:  We thank M.R.Garousi for very valuable discussions. This work is supported by the University of Guilan.
%2/16340-1389/10/14. 
\newpage
\bf\Large {Appendix\quad }}

In this appendix we want to show that the power series appearing in the energy-momentum tensors (or Lagrangians) of the nonlinear electromagnetic theories in this paper, is uniquely determined by the $SL(2,R)$ invariant structure $Tr({\cal F}^T{\cal M}_0{\cal F})$ and hence, the expression that we have found for the energy-momentum tensors would be unique. In fact, one may use $SL(2,R)$-invariant expressions $Tr({\cal F}^T{\cal M}_0{\cal F}{\cal F}^T{\cal M}_0{\cal F})$, $Tr({\cal F}^T{\cal M}_0{\cal F}{\cal F}^T{\cal M}_0{\cal F}{\cal F}^T{\cal M}_0{\cal F})$ and so on to write the energy-momentum tensors. It will be shown that these $SL(2,R)$-invariant expressions can be written in term of the power-series expansion of the $SL(2,R)$ invariant structure $Tr({\cal F}^T{\cal M}_0{\cal F})$.

Taking the trace of \reef{bisl2r}, we have
\beqa
Tr(\cF^T\cM_0\cF)&=& 2 e^{-2 \phi }Tr(F\inn F\inn F\inn F) -\frac{1}{2} e^{-2 \phi } Tr(F\inn  F)^2+e^{-3 \phi } Tr(F\inn F\inn F\inn F) Tr( F\inn F)\nonumber\\
&&-\frac{1}{4} e^{-3 \phi } Tr(F\inn  F)^3+\frac{1}{2} e^{-4 \phi } Tr(F\inn F\inn F\inn F)^2\nonumber\\
&&+\frac{1}{8} e^{-4 \phi } Tr(F\inn F\inn F\inn F) Tr(F\inn  F)^2\nonumber\\
&&-\frac{1}{16} e^{-4 \phi } Tr(F\inn  F)^4+\cO (F^{10})+\ldots\labell{FMF}
\eeqa

where we use \reef{identity} to write the $Tr(F^6)$, in terms of the $Tr(F^2)$ and $Tr(F^4)$. It could be found the following expression  for the structure $Tr(\cF^T\cM_0\cF\cF^T\cM_0\cF)$
\beqa
Tr(\cF^T\cM_0\cF\cF^T\cM_0\cF)&=& 4 e^{-2 \phi } Tr(\inn F\inn F\inn F\inn F)- e^{-2 \phi } Tr(F\inn  F)^2\nonumber\\
&&+2 e^{-3 \phi } Tr(F\inn F\inn F\inn F) Tr( F\inn F)\nonumber\\
&&-\frac{1}{2} e^{-3 \phi } Tr(F\inn  F)^3+3 e^{-4 \phi } Tr(F\inn F\inn F\inn F)^2\nonumber\\
&&-\frac{3}{4} e^{-4 \phi } Tr(F\inn F\inn F\inn F) Tr(F\inn  F)^2+\cO (F^{10})+\ldots\labell{FMFFMF}
\eeqa
where we use the following identity to write the $Tr(F^8)$, in terms of the expression including  $Tr(F^2)$ and $Tr(F^4)$.
\beqa
Tr(F\inn F\inn F\inn F\inn F\inn F\inn F\inn F)-\frac{1}{4} Tr(F\inn F\inn F\inn F)^2&-&\frac{1}{4} Tr(F\inn F\inn F\inn F) Tr(F\inn  F)^2\nonumber\\
&+&\frac{1}{16} Tr(F\inn  F)^4=0\labell{F8}
\eeqa
One can find this identity (and all similar identities that appear in this paper) by comparing the expansion of \reef{LL} with the expansion of  the BI action in the Einstein frame and explicitly check with substituting the four dimensional matrix form of the gauge fields in the identity.

Using the identity \reef{F8}, we find the following expression for $Tr({\cal F}^T{\cal M}_0{\cal F}{\cal F}^T{\cal M}_0{\cal F}{\cal F}^T{\cal M}_0{\cal F})$ 
\beqa
Tr(\cF^T\cM_0\cF\cF^T\cM_0\cF\cF^T\cM_0\cF)&=&6 e^{-4 \phi } Tr(F\inn F\inn F\inn F)^2-3 e^{-4 \phi } Tr(F\inn F\inn F\inn F) Tr(F\inn  F)^2\nonumber\\
&&+\frac{3}{8} e^{-4 \phi } Tr(F\inn  F)^4+\cO (F^{10})+\ldots\labell{FMFFMFFMF}
\eeqa
Considering the relation \reef{FMF}, \reef{FMFFMF} and \reef{FMFFMFFMF} one can write the last two structures in terms of the first structure as following
\beqa
Tr(\cF^T\cM_0\cF\cF^T\cM_0\cF)&=&2Tr(\cF^T\cM_0\cF)+\frac{1}{2}{Tr(\cF^T\cM_0\cF)}^2+\ldots\nonumber\\
Tr(\cF^T\cM_0\cF\cF^T\cM_0\cF\cF^T\cM_0\cF)&=&\frac{3}{2}{Tr(\cF^T\cM_0\cF)}^2+\frac{1}{4}{Tr(\cF^T\cM_0\cF)}^3+\ldots\nonumber
\eeqa
where we use the following identities to write the $Tr(F^{10})$ and the $Tr(F^{12})$, in terms of the expression including  $Tr(F^2)$ and $Tr(F^4)$.
\beqa
&&Tr(F^{10})+\frac{1}{64} Tr(F\inn F)^5-\frac{5}{16} Tr(F\inn F) Tr(F\inn  F\inn  F\inn  F)^2=0\nonumber\\
&&Tr(F^{12})+\frac{3}{64} Tr(F\inn F)^4Tr(F\inn  F\inn  F\inn  F)-\frac{3}{16} Tr(F\inn F)^2 Tr(F\inn  F\inn  F\inn  F)^2\nonumber\\
&&\qquad\qquad\qquad\qquad\qquad\qquad\qquad\qquad\quad -\frac{1}{16} Tr(F\inn  F\inn  F\inn  F)^3=0.\nonumber
\eeqa

%Similarly, it is not hard to investigate the structures $\Tr(\overbrace{\cF^T\cM_0\cF\cdots \cF^T\cM_0\cF}^{(4)})$,\\
% $\Tr(\overbrace{\cF^T\cM_0\cF\cdots \cF^T\cM_0\cF}^{(5)})$ and so on. However, for this purpose, one needs the identities at the higher-orders.

\end{document}